\documentstyle[11pt,flushrt,epsfig]{article}

\newcommand{\ab}{{_{{\scriptscriptstyle A,B}}}}
\newcommand{\da}{{{\delta}_{{\scriptscriptstyle A}}}}
\newcommand{\done}{{{\delta}_{{\scriptscriptstyle 1}}}}
\newcommand{\dtwo}{{{\delta}_{{\scriptscriptstyle 2}}}}
\newcommand{\doneg}{{{\delta}_{{\scriptscriptstyle 1}}^{\rm gal}}}
\newcommand{\dtwog}{{{\delta}_{{\scriptscriptstyle 2}}^{\rm gal}}}

\newcommand{\pab}{{p_{{\scriptscriptstyle AB}}}}

\newcommand{\rone}{{{\vec r}_{{\scriptscriptstyle 1}}}}
\newcommand{\ro}{\scriptscriptstyle {\rho}}
\newcommand{\rtwo}{{{\vec r}_{{\scriptscriptstyle 2}}}}

\newcommand{\rv}{{\vec r}}
\newcommand{\ra}{{{\rv}_{{\scriptscriptstyle A}}}}
\newcommand{\rb}{{{\rv}_{{\scriptscriptstyle B}}}}
\newcommand{\rh}{{\hat r}}

\newcommand{\rha}{{{\hat r}_{{\scriptscriptstyle A}}}}
\newcommand{\rhb}{{{\hat r}_{{\scriptscriptstyle B}}}}

\newcommand{\sA}{{s_{{\scriptscriptstyle A}}}}
\newcommand{\sB}{{s_{{\scriptscriptstyle B}}}}

\newcommand{\se}{{\sigma_{{\scriptscriptstyle 8}}}}
\newcommand{\va}{{{\vec v}_{{\scriptscriptstyle A}}}}
\newcommand{\vs}{{v_{{\scriptscriptstyle 12}}}}
\newcommand{\us}{{u_{{\scriptscriptstyle 12}}}}
\newcommand{\ws}{{w_{{\scriptscriptstyle 12}}}}
\newcommand{\vg}{{v_{{\scriptscriptstyle 12}}^{\rm gal}}}

\newcommand{\vu}{{{\vec u}_{{\scriptscriptstyle 12}}}}

\newcommand{\vone}{{{\vec v}_{{\scriptscriptstyle 1}}}}
\newcommand{\vtwo}{{{\vec v}_{{\scriptscriptstyle 2}}}}
\newcommand{\xb}{\bar{\xi}}

\newcommand{\xbb}{\bar{\hspace{-0.08cm}\bar{\xi}}}
\begin{document}
{\center{\Large \bf Evidence for a low density Universe \\
from the relative velocities of galaxies}\\}
{\center{\sc R. Juszkiewicz$^{* \S}$, 
P. G. Ferreira$^{* \parallel \P}$, H. A. Feldman$^{\#}$,\\}}
{\center{\sc A. H. Jaffe$^{**}$, M. Davis$^{**}$\\}}
\vskip .5in
{\small{\small
\noindent
$^{*}$~D{\'e}partement de Physique Th{\'e}orique, Universit{\'e}
de Gen{\`e}ve, CH-1211 Gen{\`e}ve, Switzerland\\
$^{\S}$~Copernicus Astronomical Center, 00-716 Warsaw, Poland\\
$^{\parallel}$~Theory Group, CERN, CH-1211, Gen{\`e}ve 23, Switzerland\\
$^{\P}$~CENTRA, Instituto Superior Tecnico, Lisboa 1096 Codex, Portugal\\
$^{\#}$~Department of Physics and Astronomy, University of Kansas,
Lawrence, KS 66045\\
$^{**}$Center for Particle Astrophysics,
University of California, Berkeley, CA94720, USA\\
}}
\vskip .3in

  {\bf The motions of galaxies can be used to
  constrain the cosmological density parameter, $\Omega$, and the
  clustering amplitude of matter on large scales. 
  The mean relative velocity of galaxy pairs, estimated from 
  the Mark III survey indicates 
  $\Omega = {\bf 0.35 _{-0.25}^{+0.35}}$.  
  If the clustering of galaxies is unbiased on large scales,
  $\Omega$ = 0.35 $\pm$ 0.15, so that an unbiased
  Einstein-de Sitter model ($\Omega =$ 1) is inconsistent with 
  the data.}


The mean relative velocity
for a pair of galaxies at positions $\rone$ and $\rtwo$ is 
$\vu = H\vec{r}$, where $\vec{r} = \rone - \rtwo$ and the
constant of proportionality $H = 100$ h km s$^{-1}$Mpc$^{-1}$ 
is the Hubble parameter {\it (1,2)}.
The quantity $0.6 <$ h $< 1$ parametrizes uncertainties 
in $H$ measurements. This law is an idealization, 
followed by real galaxies only on sufficiently large scales, 
corresponding to a uniform mass distribution. On smaller scales, 
the gravitational field induced by galaxy clusters and 
voids generates 
local deviations from the Hubble flow, called peculiar velocities. 
Correcting for this effect
gives $\vu \, = \, H\vec{r} + \vs\,\vec{r}/r$. The quantity
$\vs(r)$ is called the mean pairwise streaming velocity.
In the limit of large $r$, $\vs = 0$. In the opposite limit 
of small separations, $\us(r) = 0$ (virial equilibrium). 
Hence, at intermediate separations $\vs < 0$  
and we can expect to observe gravitational infall,
or the ``mean tendency of well-separated 
galaxies to approach each other'' {\it (3)}.
In a recent paper we derived an expression,
relating $\vs$ to cosmological parameters {\it (4)}; in another,
using Monte Carlo simulations we showed how $\vs$ can be
measured from velocity-distance surveys of galaxies {\it (5)}.
Our purpose here is to estimate $\vs(r)$ from observations and 
constrain the cosmological density parameter, $\Omega$. 

The statistic we consider was first introduced 
in the context of the 
Bogolyubov-Born-Green-Kirkwood ({\small BBGKY}) 
kinetic theory
describing the dynamical evolution of a self-gravitating
collection of particles {\it (3,6)}. 
One of the {\small BBGKY} equations is  
the so called pair conservation equation, relating
the time evolution of $\vs$ to $\xi(r)$ ---
the two-point correlation
function of spatial fluctuations in the fractional
matter density contrast {\it (3)}. Its solution 
is well approximated by {\it (4)}
\begin{eqnarray}
\vs (r) \; &=& \; - \, {\textstyle{2\over 3}} \, H\, r \,\Omega^{0.6} \,
\xbb (r)[1 + \alpha \; \xbb (r)] \;,
\label{2nd-order}\\
\xbb(r) \; &=& \; {3\, \int_0^r \, \xi(x) \, x^2 \, dx\over
r^3\,[\, 1 + \xi(r) \, ]} \; ,
\label{xb}
\end{eqnarray}
where $\, \alpha =  1.2  -  0.65  \gamma$,
$\,\gamma = - (d\ln \xi/d\ln r)_{\xi =1}\;$ and
$\Omega$ is the present density of nonrelativistic
particles. The equations above have been obtained by interpolating
between a second-order perturbative solution for $\vs(r)$ and
the nonlinear stable clustering solution. 
For a particle pair 
at separation ${\vec r}$, the streaming velocity is given by
\begin{eqnarray}
\vs(r) \, = \, \langle \, (\vone - \vtwo)
\cdot \rh \, \rangle_{\ro} \; = \;
\langle(\vone -
\vtwo )\cdot\rh \, \ws \, \rangle \; ,
\label{v12def}
\end{eqnarray}
where $\ws\, = \, (1 + \done)(1 +\dtwo )
\, [1 \; + \; \xi(r)]^{-1}$ is the pair-density weighting,
$\, \va \,$ and $\, \da \,$
are the peculiar velocity and fractional density contrast
of matter at position $\, \ra \,$, ${\small A} = 1,2 \ldots$,
the separation $r=|\rone - \rtwo|$ is fixed for all pairs, 
the hats denote unit vectors, and $\xi(r) =
\langle\done\dtwo\rangle$. The expression in square brackets 
in the definition
of $\ws$ ensures that $\langle\ws\rangle = 1$ and the pairwise 
velocity probability density integrates to unity. Note that
the pair-weighted average, $\langle\cdots\rangle_{\ro}$,
differs from simple spatial averaging, $\langle\cdots\rangle$,
by the weighting factor $\ws$. The pair-weighting makes
the average different from zero, unlike the volume average
$\langle \vone - \vtwo \rangle \, \equiv \, 0$,
which vanishes because of isotropy.

Our approximate solution of the pair conservation equation was
successfully tested against N-body simulations in 
the dynamical range $\xi \leq 10^3$ {\it (4,7)}. 
It is valid for universes filled with non-relativistic particles and it is 
insensitive to the value of the
cosmological constant {\it (2,4)}.
Eqn.(1) was derived under the
additional assumption that the probability distribution
of the initial, small-amplitude density fluctuations was Gaussian.

Until now we have also implicitly assumed that 
(i) the spatial distribution
of galaxies traces the mass distribution and that (ii) $\vs(r)$ for
the galaxies is the same as for the matter. If the galaxies are
more clustered than mass, condition (i) is broken and we have 
``clustering bias''. The galaxy two-point correlation function
is close to a power law, $\, \xi^{\rm gal}(r) \propto r^{-\gamma}$,
over three orders of magnitude in separation $r$ {\it (8)}.
This is not true for the mass correlation function $\xi(r)$ in
structure formation models of the cold dark matter ({\small CDM})
family {\it (7)}. To reconcile theory with observation, one has
to introduce a measure of bias that depends on separation and
cosmological time, $t$: $\, b^2(r,t) = \xi^{\rm gal}(r,t)/\xi(r,t)$.
Because of the pair-density weighting, clustering bias
can in principle induce ``velocity bias'' in a way similar
to systematic error propagation. This is certainly true in the 
most simplistic of all
biasing prescriptions - the ``linear biasing'', under which $b$
is a constant and, moreover, $\delta^{\rm gal} = b\delta$.
The expression for $\vg$ can be obtained from Eqn.(\ref{v12def}) 
by formally replacing the weighting function
$\ws(\done,\dtwo)$ with $\ws(\doneg,\dtwog)$.
In the linear limit, $\xi \ll 1$, we get 
$\vg = b\vs$ {\it (9)}, in qualitative agreement
with recent N-body simulations, which considered a
whole range biasing prescriptions,
allowing nonlinear and/or non-local mapping of the mass density
field onto $\delta^{\rm gal}$ 
{\it (10)}. However, there are also simulations which
show exactly the opposite: although the galaxies do not
trace the spatial distribution of mass, pairs of galaxies
behave like pairs of test particles moving in the gravitational
field of the true mass distribution, and  
$\,\vg(r) = \vs(r)\,$ {\it (11)}. 
Direct measurements of $\vg(r)$ can help us decide which 
simulations and biasing schemes are more believable than others.
Indeed, one can measure $\vg$ for different
morphological classes of galaxies. The linear bias model predicts 
$\vs^{\rm(E)}/\vs^{\rm(S)} = b^{\rm (E)}/b^{\rm (S)}$, where
the superscripts refer to elliptical (E) and spiral (S) galaxies. 
Observations suggest $b^{\rm (E)}/b^{\rm(S)} \approx 2$ and 
$b^{\rm(S)} \approx 1$ {\it (13)}. Hence, one expects
$\vs^{\rm(E)}/\vs^{\rm(S)} \approx 2$ if the linear bias model
is correct and $\vs^{\rm(E)}/\vs^{\rm(S)} = 1$
in the absence of velocity bias.

Measurements of $\vs(r)$ can be also used to determine $\Omega$. Indeed,
if the mass correlation function is well approximated by
a power law, $\xi(r) \propto r^{-\gamma}$, $\; \vs$ at a fixed
separation can be expressed in terms of $\Omega$ and the
standard normalization parameter -- $\se$. The latter
quantity is the root-mean-square contrast in the mass
found within a randomly placed sphere of radius 8 h$^{-1}$Mpc.
Unlike the conventional linear perturbative expression for 
$\vs(r) \propto \Omega^{0.6}\,\se^2\,r^{1-\gamma}$, our nonlinear
Ansatz provides the possibility of separating $\se$ from $\Omega$
by measuring $\vs$ at different values of $r$ [see the lowermost
panel in Fig.1 below; see also ref. {\it (12)}~]. 

We will now describe our measurements. The
mean difference between radial velocities of a pair of
galaxies is
$\langle \, \sA - \sB \, \rangle_{\ro} \; = \;
\vs \, \rh\cdot(\rha + \rhb)/2$,
where $\sA =  
\rha\cdot \va$ and
$\rv = \ra - \rb$. Here as before,
the latin subscripts number the galaxies
in the survey, $\; {\small A,B} = 1,2 \ldots \,$
To estimate $\vs$, we minimize the quantity
$\chi^2(\vs) \; = \; \sum_{\ab} \, \left[ (\sA - \sB) - \pab
\,\vs/2 \, \right]^2 \; \;$,
where $\pab \equiv \rh \cdot (\rha + \rhb)$ and
the sum is over all pairs at fixed separation $r = |\ra - \rb|$.
The resulting statistic is
{\it (5)}
\begin{eqnarray}
\vs (r) \; = \; {
{2 \sum \, (\sA - \sB)\, \pab }\over
{\sum \pab^2}} \;\; .
\label{estimator}
\end{eqnarray}
Monte-Carlo simulations show that this estimator
is insensitive to biases in the way galaxies are
selected from the sky and can be corrected for biases 
due to errors in the estimates of the radial distances 
to the galaxies {\it (5)}.
The survey used here is the 
Mark III standardized catalogue of galaxy peculiar
velocities {\it (14,15,16)}. 
It contains 2437 spiral galaxies with Tully-Fisher ({\small TF})
distance estimates and 544 ellipticals with $D_n-\sigma$ distances.
The total survey depth is over 120 h$^{-1}$Mpc, with homogenous
sky coverage up to 30 h$^{-1}$Mpc. The inverse 
{\small TF} and {\small IRAS} density
field corrections for inhomogeneous Malmquist bias in the spiral
sample agree with each other and give similar streaming velocities,
with lognormal distance errors of order $\sigma_{\ln d} \approx 23\%$.
For the elliptical sample, $\sigma_{\ln d} \approx 21\%$ 
and the distances assume a smooth Malmquist bias correction {\it (17)}.

The estimates from the
spiral and elliptical are remarkably consistent
with each other (Fig.1), unlike previous
comparisons using the velocity correlation tensor {\it (18,19)}.
For a velocity ratio $R = \vs^{\rm (E)}/\vs^{\rm (S)} = 1$, 
we obtain $\chi^2\simeq 1$, while 
for $R = 2$ the $\chi^2 = 2.1$.
The most straightforward interpretation
of this result is that there is no velocity bias and the
linear clustering bias model should be rejected. Its
static character and the resulting
failure to describe particle motion, induced by
gravitational instability was pointed out earlier 
on theoretical grounds {\it (20)}.  
Our results can, however,
be reconciled with linear bias model
if it is generalized to allow scale-dependence, 
$b = b(r)$. Biasing factors for both galaxy types can
be arbitrarily large at small separations, 
where $\xi(r) \gg 1$, if biasing is suppressed
at large separations, where $|\xi(r)| < 1$.
Indeed, in the nonlinear limit $\ws(b\done,b\dtwo)
\rightarrow b^2\done\dtwo/b^2\xi = \done\dtwo/\xi$, 
and hence $\vg(r) \rightarrow \vs(r)$.
 
We obtained an estimate of $\se$ and $\Omega$ from the 
shape of the $\vs(r)$ profile as follows.  
We assumed that the shape of the
mass correlation function $\xi(r)$
(but not necessarily the amplitude) is similar
the shape of the galaxy correlation function
estimated from the APM catalogue 
{\it (8)}, consistent with a power law
index $\gamma=1.75\pm0.1$ (the errors we
quote are conservative) for separations 
$\, r \le 10 \; {\rm h}^{-1}\, \mbox{Mpc} \,$. Given the
depth of the Mark III catalogue we expect the covariance between
estimates of $\vs(r)$ to be only
weakly correlated at $r < 10$
h$^{-1}$Mpc; we use N-body simulations to determine the covariance of 
the estimates over this range of scales and use a  
$\chi^2$ minimization to
obtain the 1-$\sigma$ constraints: $\se \ge 0.7$ and
$\Omega=0.35^{+0.35}_{-0.25}$. Fixing $\se = 1$ we obtain $\Omega=0.35\pm0.15$
(Fig. 2). 

We can obtain a more conservative constraint on $\se$
and $\Omega$ by examining a $\vs$ at a single separation, 
$r \equiv r_* = 10\,{\rm h}^{-1}$Mpc.
Substituting $r = r_*$ and $\xi(r) \propto r^{-1.75}$ into 
Eqns.(\ref{2nd-order})-(\ref{xb}), we get
\begin{equation}
\vs(r_*) \; =
\; - \, 605 \, {\se}^2 \Omega^{0.6}
\,(1 + 0.43\se^2)\,/(1+ 0.38{\se}^2)^2 \, {\rm km/s}
\; .
\label{10mpc}
\end{equation}
The above relation shows that at $r = r_*$, 
$\vs$ is almost entirely determined by the values of two parameters: 
$\se$ and $\Omega$. The uncertainties in the observed $\gamma$ lead to an 
error in eq.~(\ref{10mpc}) of less than $10\%$ for $\se\le 1$.
In fact, at this level of accuracy and
at this particular scale, our constraints depend only
on the value of $\Omega$ and the overall normalization $\se$
but do not depend on other model parameters, such as the
shape of $\xi(r)$. The streaming velocity, $\vs(r_*)$ 
depends on $\xi(r)$ only at $r < r_*$, so unlike
bulk flows, it is unaffected by the behavior of $\xi(r)$
at $r > r_*$ [compare our Eqn. (1) with Eqn. (21.76) in
ref. {\it (2)}~]. Moreover, the dominant contribution
to $\vs(r_*)$ comes from $\xb(r_*)$ - an average of $\xi(r)$
over a ball of radius $r_*$, so the details of the true
shape of $\xi(r)$ at $r < r_*$ have little effect on $\vs(r_*)$
as long as $\se$ (and hence, the volume-averaged $\xi$) is
fixed. Hence, 
Eqn.(\ref{10mpc}) can provide robust limits on $\se$ and $\Omega$
even if the assumption
about the proportionality of $\xi(r)$ to the {\small APM}
correlation function is dropped.
This statement can be directly tested
by comparing predictions of Eqn.(\ref{10mpc}) with predictions
od {\small CDM}-like models, all of which fail to reproduce
the pure power-law behavior of the observed galaxy correlation
function. When this test was applied to four models, recently simulated
by the Virgo Consortium {\it (7)}, we found that
for fixed values of $\se$ and $\Omega$, the predictions based
on Eqn.(\ref{10mpc}) were within $\leq 6\%$ of the $\vs(r_*)$,
obtained from the simulations {\it (21)}.

The measured value,
$-\vs (r_*) \, =
\, 280^{+68}_{-53} \;{\rm km/s}
\,$ (Fig.1), 
is inconsistent with $\se = 1$ and $\Omega=1$ at the
$99\%$ confidence level. 

Our results are compatible with a number of earlier dynamical 
estimates of the parameter $\beta \equiv \Omega^{0.6}\,  \se\,$
[$\beta$ is sometimes defined as $\Omega^{0.6}/b$, but
$\se \approx 1/b$ and the two definitions differ at most
at the 10\% level, see, for example ref.{\it (8)}~ ].
A technique, based on the
action principle {\it (22)} gives $\beta = 0.34 \pm 0.13$; 
comparisons of peculiar velocity fields with redshift surveys based on
the integral form of the continuity equation 
(called velocity-velocity comparisons) typically give
$\,\beta = 0.5-0.6 \,$ {\it (23-26)}. 
Of the velocity-velocity comparisons, the one with the smallest
error bars is the  {\small VELMOD} 
estimate: $\;\beta=0.5\pm0.05$ {\it (24)}. 
This constraint has several advantages
over others; in particular,
it correctly takes into account cross calibration errors between 
different Mark III subcatalogues.
To illustrate the consistency of our results with velocity-velocity
studies we will now compare our limits on $\se$ and $\Omega$,
derived from the shape of $\vs(r)$ for a range of separations
with constraints from
our measurement of $\vs(10 \, {\rm h}^{-1}\;\mbox{Mpc})$ alone,
combined with limits from {\small VELMOD} (Fig.2).
Again
we find that a low-$\Omega$ universe is favoured: $\Omega < 0.65$ and
$\se > 0.7$. 
The concordance region overlaps with the
constraint derived from our measurements of
$\vs(r)$. 

Our results disagree with the {\small IRAS-POTENT} estimate,
$\beta = 0.89 \pm 0.12$ {\it (27)}. 
The {\small IRAS-POTENT} analysis is based on the continuity
equation in its differential form; it uses a rather
complicated reconstruction technique to recover the full
velocity field from its radial component. The reason for the
disagreement is not clear at present. One can think of 
at least two
possible sources of systematic errors in the {\small IRAS-POTENT}
analysis: 
(i) the reconstruction scheme itself (e.g., taking
spatial derivatives of noisy data), and (ii) the nonlinear
corrections adopted. 
The nonlinear corrections diverge like $\, \Omega^{-1.8}$ 
in the limit $\Omega \rightarrow 0$ {\it (27)}. 
By contrast, the accuracy of the 
nonlinear corrections for the velocity-velocity is insensitive   
to $\Omega$. The velocity-velocity approach is also
simpler than {\small IRAS-POTENT} because it
does not involve the reconstruction of the full velocity vector
from its radial component measurements (although both approaches do
require a reconstruction of galaxy positions from their redshifts). 
Note that our method 
is direct, not inverse: it does not require any reconstruction at all. 

Finally, there is a potential
caveat in the ``no velocity bias'' 
assumption in our own analysis.
Although this assumption is based on
empirical evidence from the two sets of galaxy types,
the observational data is noisy and involves
non-trivial corrections for Malmquist bias which
could affect the two samples differently. Application
of our approach to different data sets will clarify
these issues. If contrary to our preliminary results,
the streaming velocity turns out to be
subjected to bias after all, such a finding
may affect our estimates of $\se$ and $\Omega$,
based on the shape of the $\vs(r)$ profile but
not our rejection of the unbiased
Einstein-de Sitter model. In this sense our differences
with the {\small IRAS-POTENT} analysis do not depend
on the presence or absence of velocity bias.

The advantages of the new statistic we have used here
can be summarized as follows.
First,  $\vs$ can be estimated directly from velocity-distance
surveys, without subjecting the observational data to multiple
operations of spatial smoothing, integration and differentiation, 
used in various reconstruction schemes. Second, unlike
cosmological parameter estimators based on the acoustic peaks, expected
to appear in the cosmic microwave background power spectrum {\it (28)},
the $\Omega$ estimate based on $\vs$ is model-independent. Finally,
our approach offers the possibility to break the degeneracy between
$\Omega$ and $\se$ by measuring the $\vs(r)$ at different
separations.

\noindent
{\Large {\underline {References and Notes}}}

\small
\noindent
(1) E.P. Hubble, 
{\it Proc. Natl. Acad. Sci. USA} {\bf 15}, 168 (1929).\\ 
\noindent
(2) P.J.E. Peebles, {\it Principles
of Physical Cosmology}, pp. 340-342 
(Princeton University Press, Princeton, 1993).\\
\noindent
(3) P.J.E. Peebles, {\it The
Large--Scale Structure of the Universe}, pp. 266-268
(Princeton University Press, Princeton, 1980).\\
\noindent
(4) R. Juszkiewicz, V. Springel, R. Durrer,  
{\it Astrophys. J.} {\bf 518}, L25 (1999).\\
\noindent
(5) P.G. Ferreira, R. Juszkiewicz, H. Feldman, 
M. Davis, A.H. Jaffe, 
{\it Astrophys. J.} {\bf 515}, L1 (1999).\\
\noindent 
(6) M. Davis, M., P.J.E. Peebles,
{\it Astrophys. J. Suppl.} {\bf 34}, 425 (1977).\\
\noindent
(7) A. Jenkins {\it et al.}, {\it Astrophys. J.}
{\bf 499}, 20 (1998).\\
(8) G. Efstathiou, in 
{\it Les Houches, Session LX. Cosmology and large scale structure.}
(eds. R. Schaeffer, {\it et al.}) 
107-252 (North-Holland, Amsterdam, 1996).\\ 
(9) K.B. Fisher, M. Davis,
M. Strauss, A. Yahil, J. Huchra,
{\it Mon. Not. R. Astron. Soc.} {\bf 267}, 927 (1994).\\ 
(10) V.K. Narayanan, A.A. Berlind, D.H. Weinberg,
{\tt astro-ph/9812002} (1998).\\
(11) G. Kauffmann, J.M. Colberg,
A. Diaferio, S.D.M. White,
{\it Mon. Not. R. Astron. Soc.} {\bf 303}, 188 (1999).
To linear bias enthusiasts, this result may appear puzzling. 
A textbook example of such a possibility is
the spatial distribution of the gas and stars in a galaxy.
Their distribution itself is biased, while their relative
velocities at separations comparable to the radius
of the dark halo are not.\\
(12) N-body simulations show that the stable clustering 
solution ($\us = 0, -\vs(r)/Hr = 1$)
occurs for $\xi > 200$ [ref.{\it (7)},
see also B. Jain,
{\it Mon. Not. R. Astron. Soc.} {\bf 287}, 687 (1997)].
Note that in the limit $r \rightarrow 0$ our Ansatz
gives $-\vs/Hr \rightarrow [2/(3-\gamma)]\Omega^{0.6}(1+\alpha)
$, which for the range of parameters considered is
generally different from unity (although of the
right order of magnitude). It is possible
to improve our approximation and satisfy the
small separation boundary condition by
replacing $(2/3)\Omega^{0.6}$ with 
the expression $[(1-\gamma/3)F\xi(r) + (2/3)\Omega^{0.6}]
/[F\xi(r) +1]$, where $F \approx 1/100$ is a fudge
factor which ensures that $-\vs/Hr \rightarrow  1$ when
$\xi \gg 100$ while the perturbative, large-scale
solution remains unaffected. However, the stable clustering
occurs at separations smaller than $200$h$^{-1}$kpc,
which is a tenth of the smallest
separation we consider here. As a result, in
our range of separations $r$,
Eqn.(1) is as close to fully nonlinear N-body
simulations as the improvement, suggested above
[see ref. {\it (4)}, Fig. 2].  
\\
(13) M.A. Strauss, J.A. Willick,   
{\it Phys. Rep.} {\bf 261}, 271 (1995).\\
(14) J.A. Willick, S. Courteau, S.M. Faber, D. Burstein,
A. Dekel, 
{\it Astrophys. J.} {\bf 446}, 12-38 (1995).\\ 
(15) J.A. Willick, {\it et al.}, 
{\it Astrophys. J.} {\bf 457}, 460 (1996).\\ 
(16) J.A. Willick,  {\it et al.},  
{\it Astrophys. J. Suppl.} {\bf 109}, 333 (1997).\\
(17) D. Lynden-Bell, {\it et al.}, 
{\it Astrophys. J} {\bf 329}, 19 (1988).\\
(18) K.M. G\'orski, M. Davis, M.A. Strauss, 
S.D.M. White, A. Yahil, 
{\it Astrophys. J.} {\bf 344}, 1 (1989).\\ 
(19) E.J. Groth, R. Juszkiewicz, J.P. Ostriker,  
{\it Astrophys. J.} {\bf 346}, 558 (1989). \\
(20) Under realistic circumstances one expects that gravitational
growth of clustering pulls the mass with the galaxies. As a result, 
any clustering bias, introduced at the epoch 
of galaxy formation is likely to evolve towards unity at late times,
at variance with the linear bias model, where $b$ is time-independent
[J. N. Fry, {\it Astrophys. J.} {\bf 461}, L65 
(1996); P. J. E. Peebles, {\tt astro-ph/}9910234]. 
A similar behavior was seen in N-body
simulations {\it (11)}. 
Note also that the linear bias relation
$\delta^{\rm gal} = b\delta$ is a conjecture which
generally does not follow from the
relationship between the correlation
functions $\xi^{\rm gal} = b^2\xi$ unless we 
restrict our model to a narrow subclass of random fields.\\
(21) Imagine that one of the
{\small CDM}-like models is a valid description of our Universe.
Let us choose the so-called {\small $\Lambda$CDM} model,
recently simulated by the Virgo Consortium {\it (7)}.
It is defined by parameters
$h = 0.7, \, \Omega = 0.3, \, \se = 0.9$ and $\Omega_{\Lambda}
= 0.7$, which is the cosmological constant's contribution to
the density parameter. This model requires
scale-dependent biasing because the predicted
shape of $\xi(r)$ differs widely
from the observed galaxy correlation function: at separations
${\rm h}r/{\rm Mpc} = 10, \, 4, \, 2$ and 0.1 the logarithmic slope
of the mass correlation function reaches the values,
given by $\gamma = 1.7, \, 1.5, \, 2.5$ and 1, respectively.
As we expected, however, these wild oscillations do not
affect the resulting $\vs(r_*)$. Indeed, the N-body simulations
{\it (7)} give $\vs(r_*) = - 220$ km/s, while 
substituting $\se = 0.9$ and $\Omega = 0.3$ in Eqn.(\ref{10mpc})
gives $\vs(r_*) = - 225$ km/s.\\ 
(22) E.J. Shaya, P.J.E. Peebles, R.B. Tully,
 {\it Astrophys. J.} {\bf 454}, 15 (1995).\\
(23) M. Davis, A. Nusser, J.A. Willick, 
{\it Astrophys. J.} {\bf 473}, 22 (1996).\\
(24) J.A. Willick, M.A. Strauss, A. Dekel, 
T. Kolatt, 
{\it Astrophys. J.} {\bf 486}, 629 (1997).\\ 
(25) L.N. da Costa,  {\it et al.},
{\it Mon. Not. R. Astron. Soc.} {\bf 299},
425 (1998).\\
(26) J. A. Willick, M.A. Strauss, 
{\it Astrophys. J.} {\bf 507}, 64 (1998).\\ 
(27) Y. Sigad, A. Eldar, A. Dekel, M.A. Strauss, 
A. Yahil, {\it Astrophys. J. Suppl.}
{\bf 109}, 516 (1997).\\ 
(28) A.G. Doroshkevich, Ya.B. Zeldovich, R.A. Sunyaev,
{\it Sov. Astron.} {\bf 22}, 523 (1978).\\ 
(29) We are grateful to Luiz da Costa
for encouragement at early stages of this project. We also
benefitted from discussions with
Stephane Courteau, Ruth Durrer, Kris G\'orski 
and Jim Peebles. RJ was supported by the KBN
(Polish Government), the Tomalla Foundation
(Switzerland), and the Poland-US M. Sk{\l}odowska-Curie
Fund. MD and AHJ were supported by grants from the
National Science Foundation and NASA. HAF and RJ were
supported by the NSF-EPSCor and GRF.
We thank the Organizers of Summer 1997 workshop 
at the Aspen Center for Physics, where this work began.\\
(30) Correspondence should be 
addressed to PGF (e-mail: {\tt pgf@astro.ox.ac.uk}).

\eject

\noindent
\vskip .25in
\noindent
{\bf Figure 1:} The streaming velocities of 2437 spiral galaxies
(top panel) and 544 elliptical galaxies (center panel) estimated 
from the Mark III catalogue. The error bars are the
estimated 1$\sigma$ uncertainties in the measurement due to
lognormal distance errors, sparse sampling (shot noise) and
finite volume of the sample (sample variance). 
The error bars 
were estimated from mock catalogues described in ref. {\it (5)}.
The small
sample volume also introduces correlations between
measurements of $\vs(r)$ at different values of $r$. 
To guide the eye,
and show that although the two samples have different noise levels
(because of much smaller number of galaxies in the elliptical
sample), the $\vs(r)$ signal in both cases is similar, we
also plot $\vs(r)$ calculated from equation (1)
for a $\xi \propto r^{-1.75}$ power-law model with
$\se = 1.25$ and $\Omega = 0.3$. Three theoretical $\vs(r)$
curves are plotted (bottom panel) with $\xi \propto r^{-1.75}$,
$\se\Omega^{0.6}=0.7$ and $\se=0.5$ (solid line), $1$
(dotted line) and $1.5$ (dashed line). These curves show
how measurements of $\vs(r)$ can break the degeneracy between
$\Omega$ and $\se$.

\vskip .2in
\noindent
{\bf Figure 2}: The blue region constrains the viable values of $\Omega$, the
fractional mass density of the universe, and $\se$, the variance of
mass fluctuations at $r = 8\, h^{-1}$Mpc, from the combination of the
constraints on the streaming velocities (red region) and $\beta= \,\se\,
\Omega^{0.6}$ (green region). The streaming velocities are constrained at
$r = 10\, h^{-1}$Mpc from the Mark III catalogue of peculiar velocities
and $\beta$ is measured using the {\sc VELMOD} comparison between the
Mark III catalogue and the velocity field inferred from the {\sc IRAS}
redshift survey. The dashed line defines the 1-$\sigma$ obtained from
comparing expressions 1 and 2 with the Mark III catalogue from $2$h$^{-1}$Mpc
to $10$h$^{-1}$ Mpc.
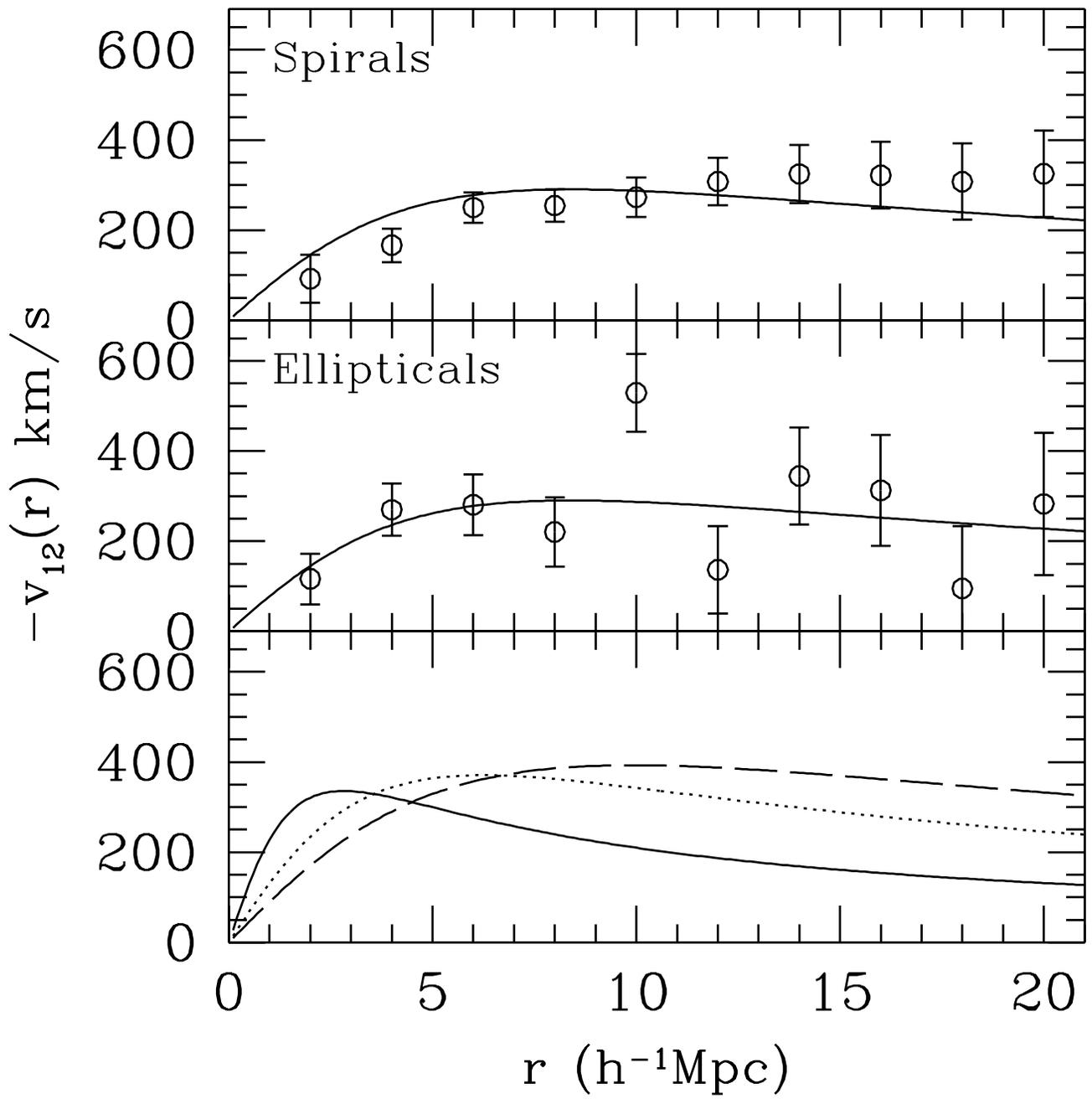
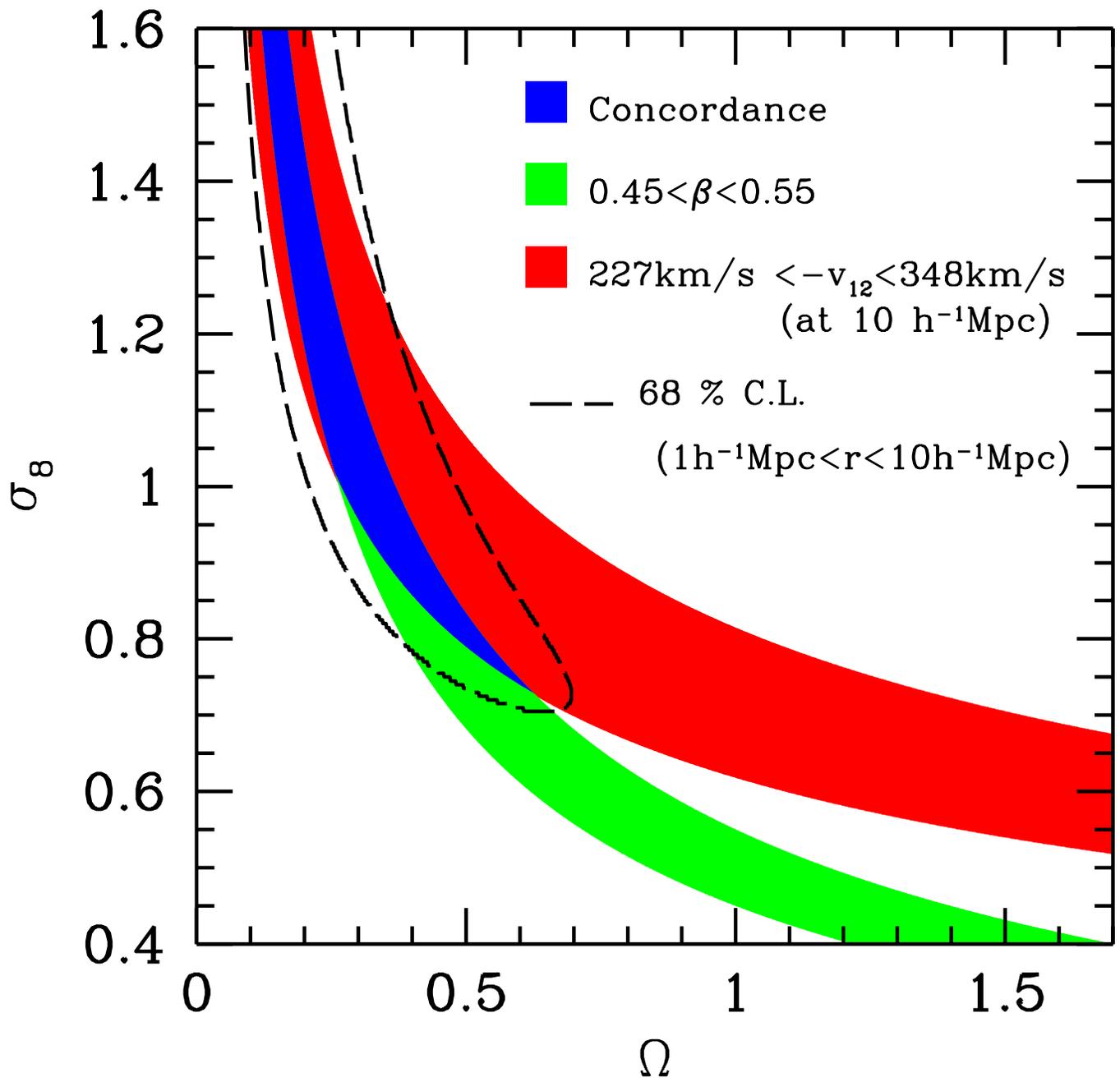
\end{document}